\documentclass[twocolumn,aps,prb,superscriptaddress,address,amsmath,amssymb,showpacs]{revtex4-1}
\usepackage[latin9]{inputenc}
\usepackage{amsmath}
\usepackage{amssymb}
\usepackage{graphicx}
\usepackage{esint}
\usepackage[unicode=true,pdfusetitle,
 bookmarks=true,bookmarksnumbered=false,bookmarksopen=false,
 breaklinks=false,pdfborder={0 0 1},backref=false,colorlinks=false]
 {hyperref}



\begin{document}

\title{Pure spin current devices based on ferromagnetic topological insulators}

\author{Matthias Götte, Michael Joppe and Thomas Dahm}

\affiliation{Universität Bielefeld, Fakultät für Physik, Postfach 100131, D-33501
Bielefeld, Germany}

\date{\today}

\begin{abstract}
Two-dimensional topological insulators possess two counter propagating
edge channels with opposite spin direction. Recent experimental progress
allowed to create ferromagnetic topological insulators realizing a
quantum anomalous Hall (QAH) state. In the QAH state one of the two
edge channels disappears due to the strong ferromagnetic exchange
field. We investigate heterostuctures of topological insulators and
ferromagnetic topological insulators by means of numerical transport
calculations. We show that spin current flow in such heterostructures
can be controlled with high fidelity. Specifically, we propose spintronic
devices that are capable of creating, switching and detecting pure
spin currents using the same technology. In these devices electrical
currents are directly converted into spin currents, allowing a high
conversion efficiency. Energy independent transport properties in
combination with large bulk gaps in some topological insulator materials
may allow operation even at room temperature. 
\end{abstract}

\maketitle
In contrast to charge based electronic devices, spintronic devices
are supposed to include or solely use the spin degree of freedom of
charge carriers \cite{Wolf,Zutic}. This requires materials and methods
that allow the creation and control of pure spin currents in solid-state
systems. Currently there exist a few methods capable of creating pure
spin currents, like spin Hall effect\cite{Hirsch,Hoffmann}, spin
pumping\cite{Mosendz,Czeschka,Weiler-PRL2013} and spin Seebeck effect\cite{Weiler-PRL2013,Uchida,Weiler-PRL2012}.
The detection of spin currents usually exploits the inverse spin Hall
effect\cite{Czeschka,Mosendz,Weiler-PRL2013,Uchida,Weiler-PRL2012}.
However, the efficiency of these methods is low regarding the power
needed to create a sizeable spin current\cite{Weiler-PRL2013}.

Topological insulators (TIs) are materials, which are insulating in
the bulk but possess conducting states at the surface or, in the two-dimensional
(2D) limit, at the edges\cite{Bernevig,Fu}. The spin of these edge
states is locked with the propagation direction along the edge, i.e.
electrons with opposite spin orientation move in opposite direction\cite{Ando,Hsieh,Bruene,Pan:PRL106}.
In addition, the surface states are topologically protected which
precludes backscattering and conserves the spin-momentum locking of
the edge states \cite{Sheng}. These properties make TIs promising
candidates for spintronic devices\cite{Hasan,Pesin}.

Recent experimental progress has resulted in creation of ferromagnetic
topological insulators. Ferromagnetism is either induced by doping
with transition metal atoms \cite{Hor2010,Xu2012,QAHScience,Checkelsky,Kou2014,QAHVdoped}
or by the proximity effect \cite{Vobornik,Wei,Yang2014,Lang2014,Li2015,LiPRB15}.
When the ferromagnetic exchange field (FEF) is directed perpendicular
to the 2D TI sheet the edge state dispersion remains robust \cite{Paananen-PRB2013,PGGD}
and does not acquire a gap. When the FEF exceeds a critical strength,
one pair of edge states is pushed into the bulk and disappears resulting
in a quantum anomalous Hall state (QAH) in which only a single spin
direction can propagate along the edge in a single direction, similar
as in the quantum Hall state \cite{CXLiu2008,QAHtheory}. The QAH
state has been demonstrated experimentally in several ferromagnetic
topological insulators \cite{QAHScience,Checkelsky,Kou2014,QAHVdoped}.
Such ferromagnetic TIs thus allow to switch and selectively turn off
certain edge state channels by changing the magnetization direction.
Because of a large bulk gap in some TI materials \cite{Pauly}, e.g.
about $0.3$ eV in Bi$_{2}$Se$_{3}$ \cite{Zhang:NPhys09,Yazyev}
and even larger gaps in newly predicted 2D materials\cite{Song,Luo},
devices based on these materials could even operate at room temperature
\cite{Lang2014}.

Recent experiments and proposals for TI spintronic devices mainly
focus on the injection of spin or spin polarized electrons into the
surface states of TIs \cite{Goette,Shiomi,Yokoyama,Tian} or on the
manipulation of spin polarized currents \cite{Krueckl,Michetti} but
do not consider pure spin currents.

In the present work, we demonstrate by numerical transport calculations
that heterostructures combining topological insulators and ferromagnetic
topological insulators can steer spin currents with high fidelity
in a controlled way. We present devices that allow the creation, switching
and detection of pure spin currents, all using the same type of heterostructures.
An electrical voltage is directly converted into a pure spin current,
which allows a higher conversion efficency than the above mentioned
known methods for spin current generation and detection. Our proposals
are supported by numerical time evolution of wave packets on finite
2D lattices. We are using a model and parameters suitable for Bi$_{2}$Se$_{3}$
thin films (see methods section), however our proposals can be applied
to other materials as well.

\section*{Results}

\subsection*{Gapless states}

\begin{figure}
\includegraphics[width=1\columnwidth]{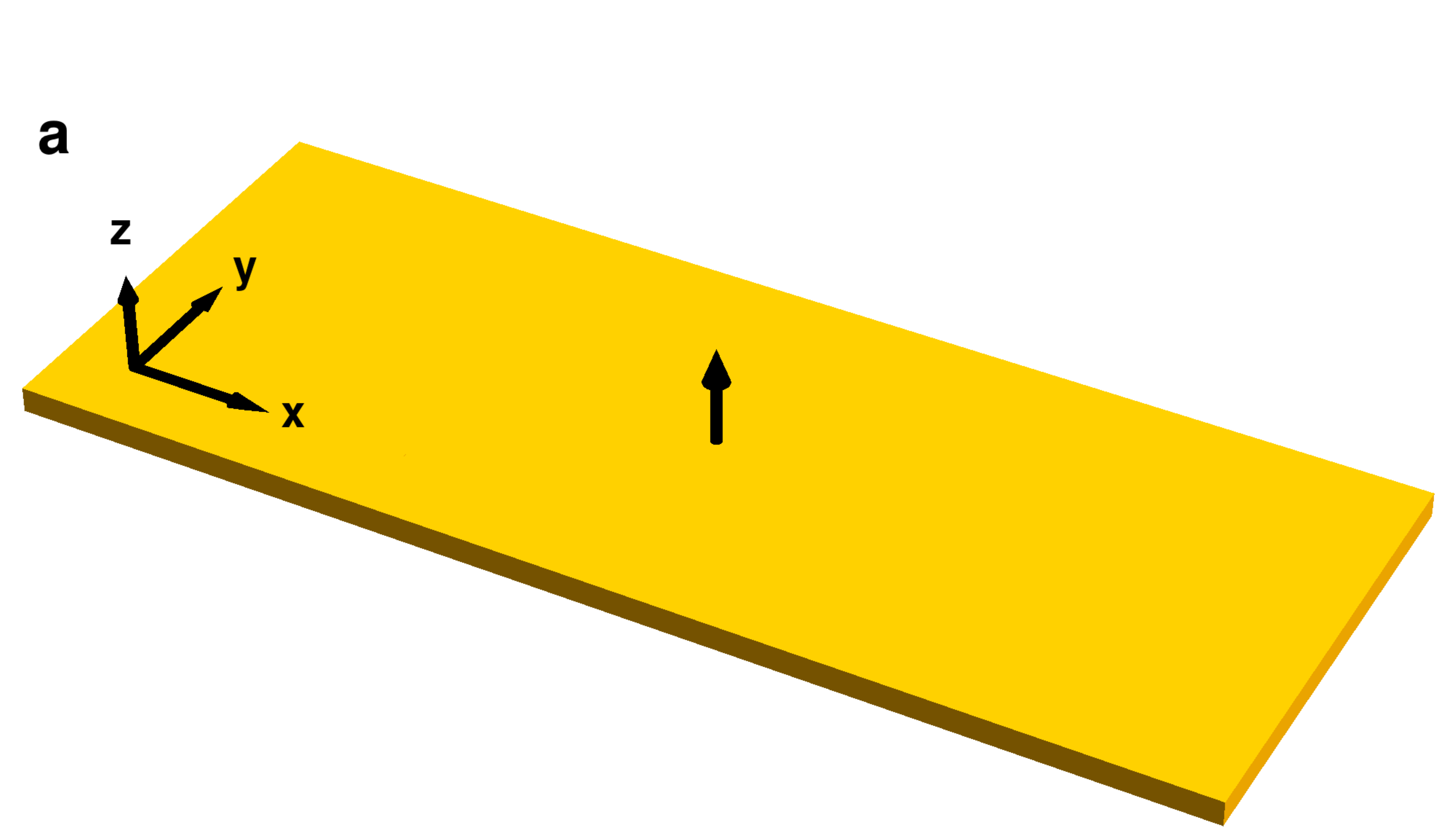}

\includegraphics[width=1\columnwidth]{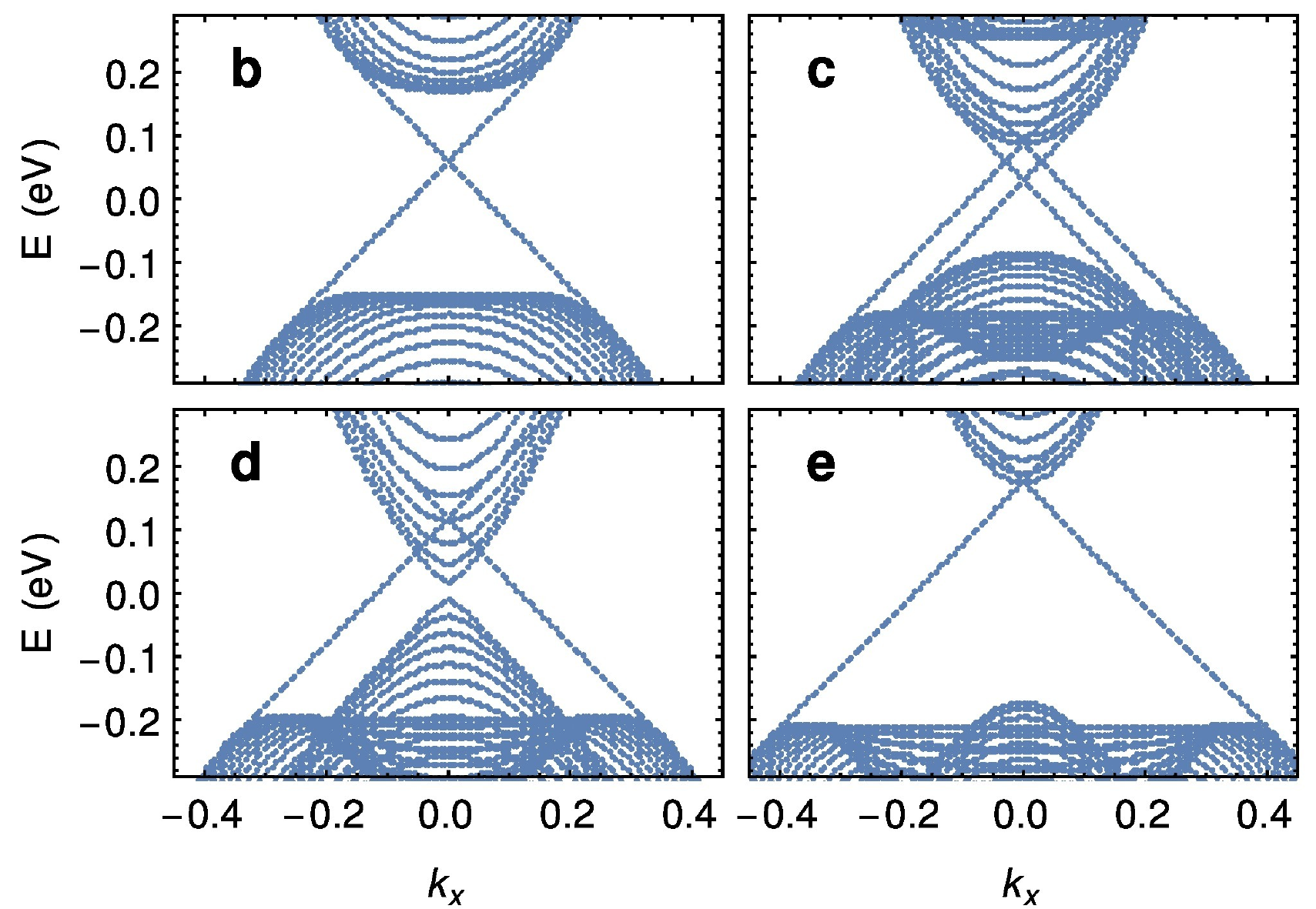}

\caption{\textbf{\label{fig:Dispersion}Dispersion.} Numerical dispersion of
bulk and edge states of a TI bounded by vacuum in $y$-direction as
shown in (\textbf{a}). Different strengths $V_{z}$ of an FEF applied
in $z$-direction are shown. (\textbf{b}) $V_{z}=0$. Edge states
are twofold degenerate, with counterpropagating spin-up and spin-down
states at each edge. (\textbf{c}) $V_{z}=0.5M_{\textrm{2D}}$, with
$M_{\textrm{2D}}=0.17\,\textrm{eV}$ (see methods section). Spin-up
and spin-down edge states split. The bulk gap becomes smaller. (\textbf{d})
$V_{z}=M_{\textrm{2D}}$. Spin-down edge states are completely removed
and the bulk gap is nearly closed. (\textbf{e}) $V_{z}=2M_{\textrm{2D}}$.
The bulk gap has reopened to approximately the same energy range as
in (\textbf{b}) without FEF. Only non-degenerate spin-up edge states
are present. The system is now in a quantum anomalous Hall state. }
\end{figure}

The goal of this work is to construct spintronic devices from TIs
by using ferromagnetic exchange fields (FEFs). Before we examine 
the above mentioned heterostructures,
we first need to understand how the gapless edge states behave in
the presence of FEFs. For that purpose we first review the transition of the 
Quantum Spin Hall (QSH) state into a QAH state with increasing 
FEF following ref. ~\onlinecite{Paananen-PRB2013}.
We have calculated the dispersion of edge and
bulk states by exact diagonalization of a lattice model for a thin ferromagnetic
Bi$_{2}$Se$_{3}$ strip as detailed in the methods section below.
Results are shown in Fig.~\ref{fig:Dispersion} and discussed in
the following for different strengths $V_{z}$ of an FEF applied perpendicular
to the surface plane.

In a pure 2D TI, gapless states exist only at the edges of the TI
where it is in contact to an ordinary insulator or vacuum (Fig.~\ref{fig:Dispersion}b).
They are twofold degenerate at each edge (fourfold degenerate, considering
both edges as in Fig.~\ref{fig:Dispersion}) with spin polarization
in $z$-direction (perpendicular to the surface plane of the 2D sheet)
and have an approximately linear dispersion. Note, however that
in general each edge state is a superposition of different orbital
states with different spin components, which partially compensate
each other. Therefore it is more appropriate to speak of
two orthogonal pseudo-spin states rather than spin-states.
Whereas the absolute
value of the spin polarization of both states is the same, one of
them is dominated by spin-up and propagates clockwise along the edges
while the other one is dominated by spin-down and propagates counterclockwise.
In the following we will refer to the first of these pseudo-spin
states as spin-up and
to the second ones as spin-down and indicate them in any figure by
green and red color, respectively.

An FEF in $z$-direction lifts the degeneracy of the edge states without
opening a gap (Fig.~\ref{fig:Dispersion}c-e) \cite{RLChu2}. For
small FEF, the edge state dispersions are shifted in momentum (Fig.~\ref{fig:Dispersion}c).
When the FEF becomes of the order of half the bulk gap, the bulk bands
touch closing the bulk gap and absorbing one pair of edge states into
the bulk (Fig.~\ref{fig:Dispersion}d). When the FEF becomes larger
than half of the initial bulk gap, the bulk gap reopens leaving only
a single pair of edge states. If the field is strong enough ($\left|V_{z}\right|\gtrsim0.34\,\textrm{eV}$
in our case, see Fig.~\ref{fig:Dispersion}e), one pair of edge states
is completely removed and the bulk gap is restored. The system is
then in a quantum anomalous Hall state \cite{Paananen-PRB2013,PGGD}.
The remaining states are only shifted in momentum and have a slightly
modified group velocity. The degree of spin polarization remains unchanged,
however. Which states remain inside the bulk gap depends on the sign
of the FEF, in which a positive (negative) FEF removes spin-down (up)
states. This analysis of the influence of an FEF on the edge states
shows that one can selectively remove one pair of edge states by applying
a moderate exchange field of the order of 0.3 eV. In the following
we will refer to this situation of a TI with a sufficiently strong
FEF as ferromagnetic topological insulator (FTI).

\begin{figure*}
\begin{minipage}[c][1\totalheight][t]{0.33\textwidth}%
\includegraphics[width=1\textwidth]{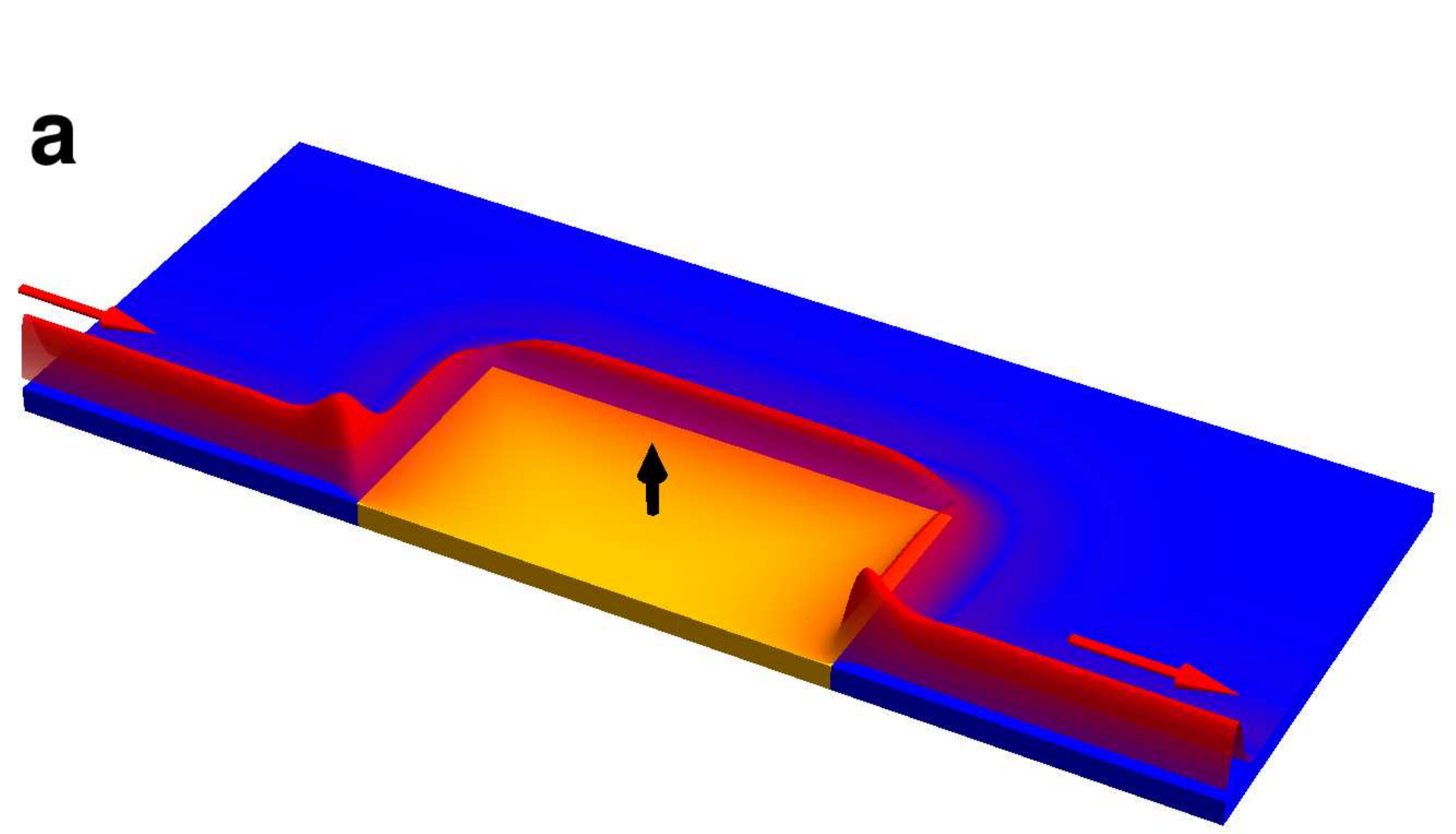}

\includegraphics[width=1\textwidth]{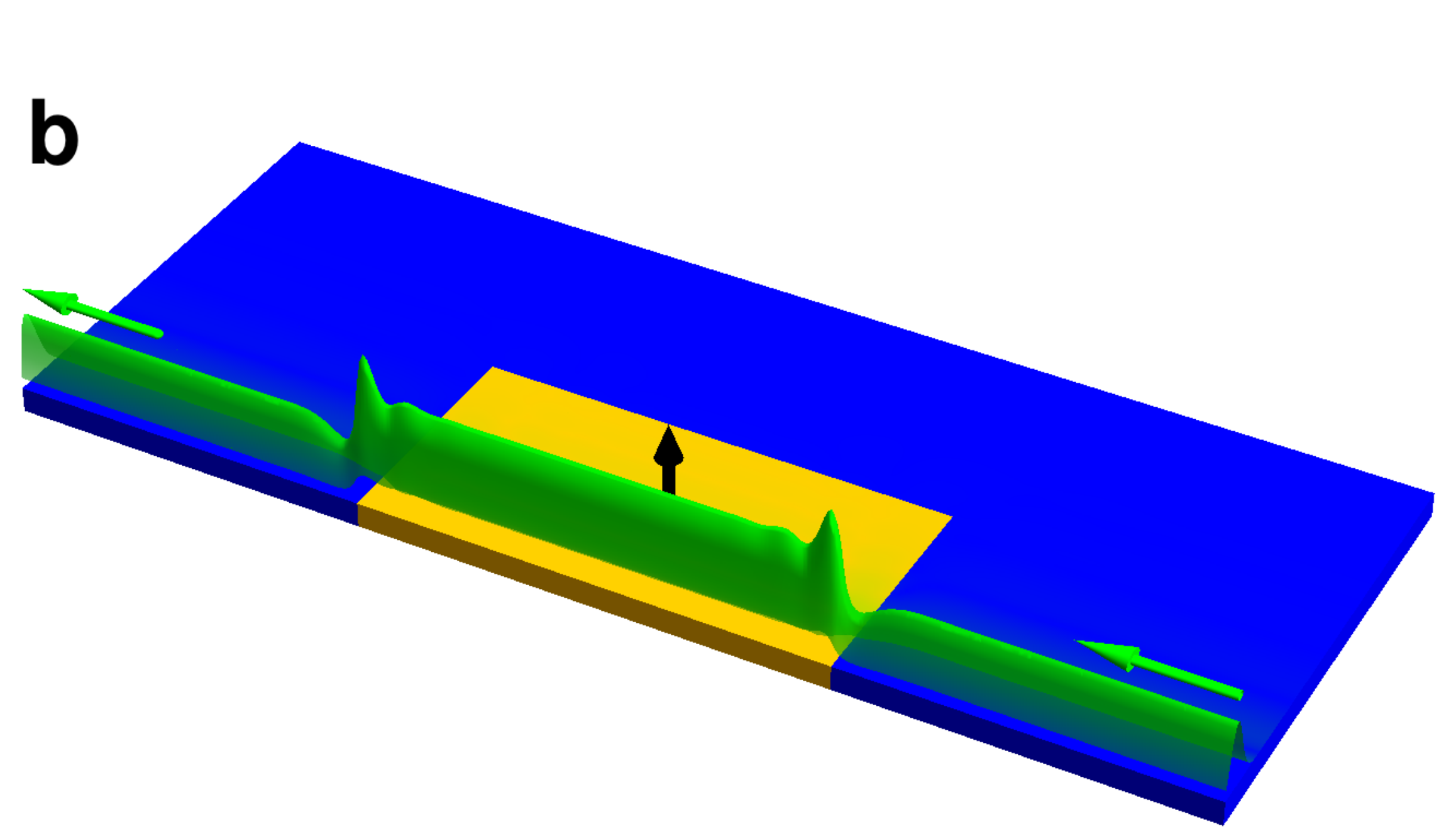}%
\end{minipage}%
\begin{minipage}[c][1\totalheight][t]{0.33\textwidth}%
\includegraphics[width=1\textwidth]{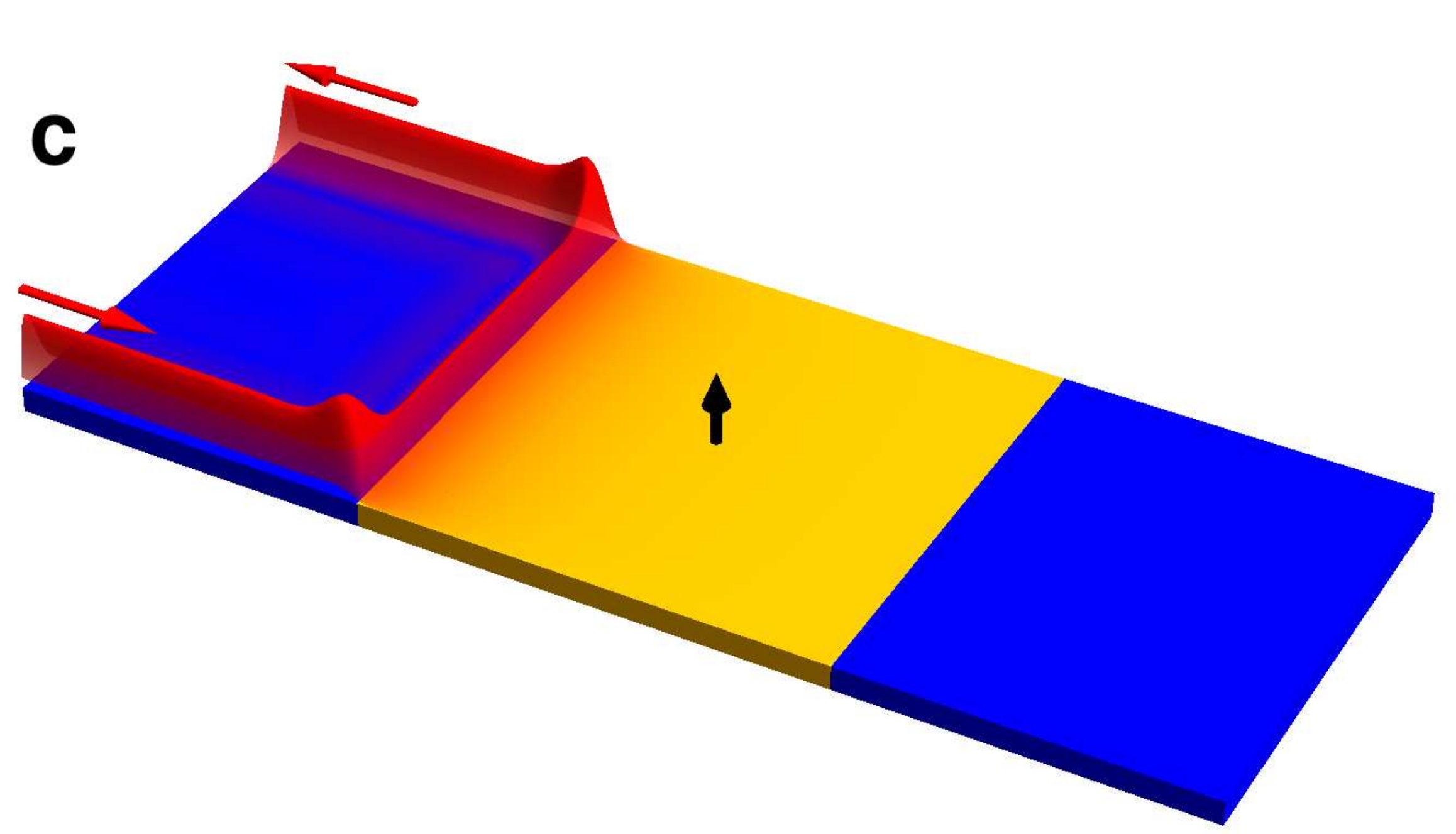}

\includegraphics[width=1\textwidth]{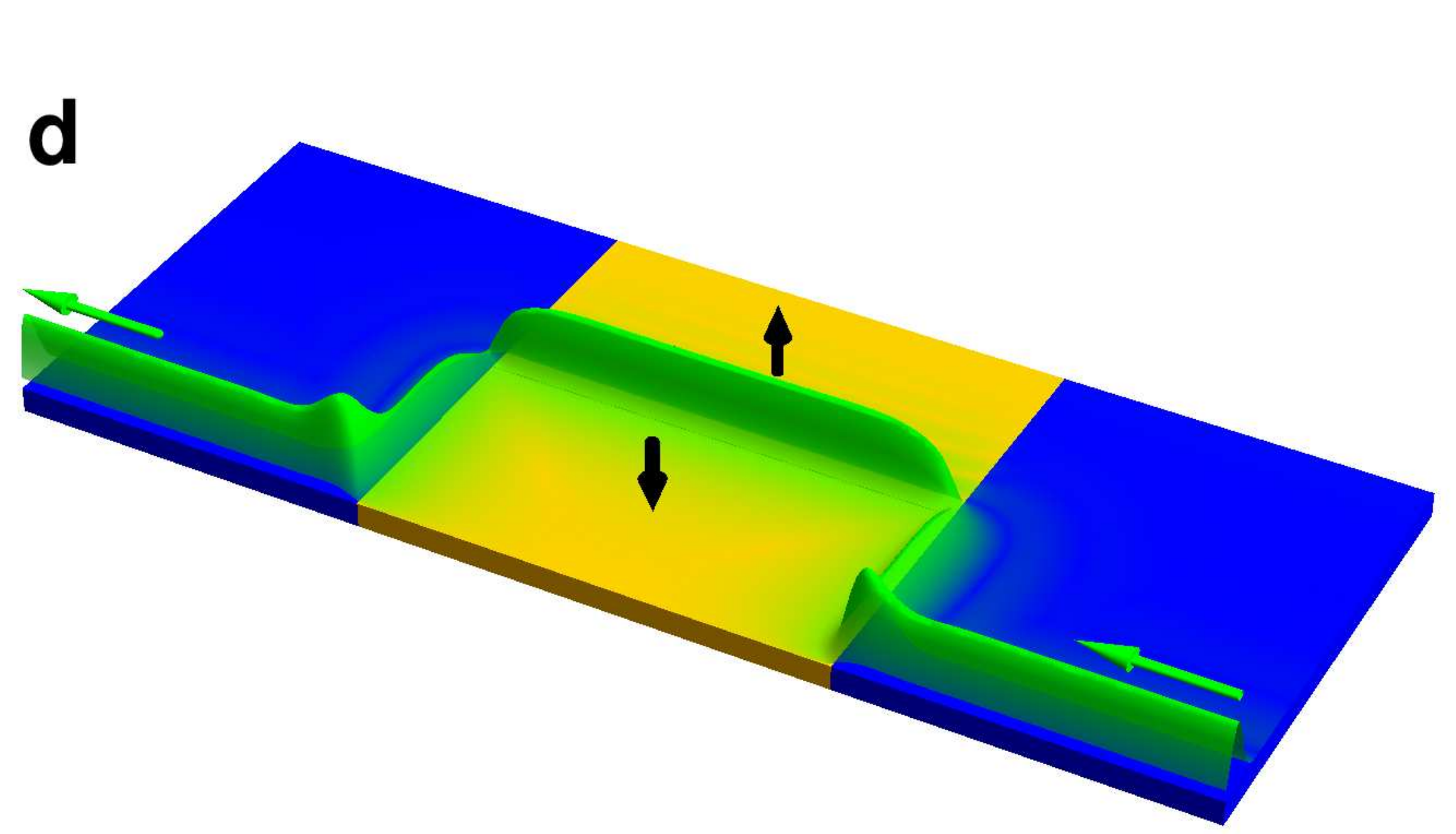}%
\end{minipage}%
\begin{minipage}[c][1\totalheight][t]{0.33\textwidth}%
\includegraphics[width=1\textwidth]{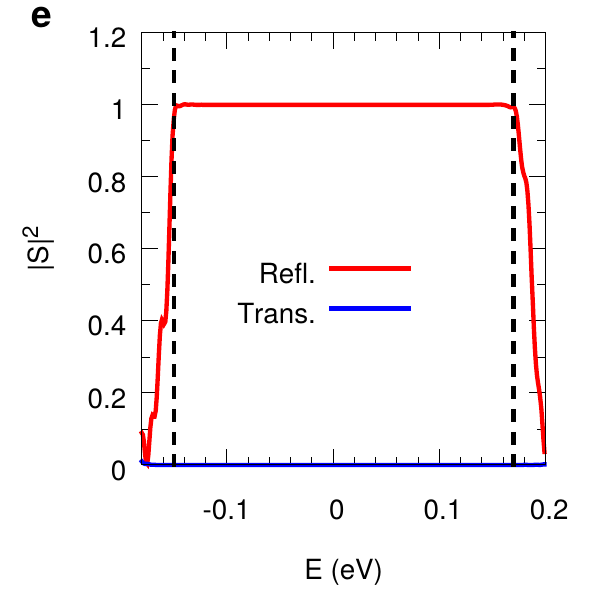}%
\end{minipage}

\caption{\label{fig:DoS}\textbf{Propagation of gapless states at local FEFs.}
(\textbf{a})-(\textbf{d}) Local density of states at different local
FEFs (orange; black arrows indicate the polarization) on a narrow
TI strip (blue). Green color indicates spin-up states and red color
spin-down states. The corresponding red and green arrows show the
entrance and exit of an electron in that state. At the lower edge,
spin-up states come from the right and spin-down states from the left.
Fermi energy is choosen as $E=0$, approximately in the center of
the bulk gap. (\textbf{e}) Energy dependent reflection and transmission
probabilities corresponding to \textbf{c}. Reflection is nearly perfect
over the largest part of the TI bulk gap. The gap edges are shown
by the vertical dashed lines. 
Setups \textbf{a}, \textbf{b} and \textbf{d} show perfect
transmission.}
\end{figure*}

As a next step, we consider heterostructures of a TI without FEF and
an FTI as shown in Fig.~\ref{fig:DoS}. In order to determine the
behavior of the edge states in such an inhomogeneous situation, we
have done numerical quantum transport calculations \cite{Krueckl}
as detailed in the methods section. We prepare initial electron wave
packets in different edge channels and follow their evolution through
the system by solving the time dependent Schroedinger equation. The
path an electron takes is shown in red color for spin-down states
and in green color for spin-up states. Our results show that the presence
of an FTI area in a TI does not destroy the pair of gapless states
discussed above, but instead pushes them away from the TI edge towards
the interface of the TI with the FTI area. For an FTI with positive
polarization a spin-down electron coming from the left cannot propagate
inside the FTI area and thus takes a detour around the FTI area and
then moves on at the same edge of the TI (Fig.~\ref{fig:DoS}a).
On the other hand, a spin-up electron coming from the right stays
at the edge and moves straight through the FTI (Fig.~\ref{fig:DoS}b).
Changing the polarization of the FTI interchanges the paths of the
counterpropagating electrons, i.e. the electron from the right takes
the detour and the one from the left moves straight through. When
the FTI area is extended to the opposite edge of the TI, as shown
in Fig.~\ref{fig:DoS}c, one of the spin states is removed at both
edges, i.e. for an FTI with positive polarization, an incoming spin-down
electron can no longer pass from left to right and vice versa. As
the FEF in $z$-direction conserves the pseudospin
of the electron, it cannot be reflected back
at the same edge but instead moves along the FTI-TI interface towards
the other TI edge where it propagates back. On the other hand an incoming
spin-up electron starting at the upper left corner can propagate at
the upper edge right through the FTI area. In this way the structure
shown in Fig.~\ref{fig:DoS}c can be used to selectively block one
of the two edge channels from propagating from left to right. The
behavior at the interface of two FTIs with opposite polarization (Fig.~\ref{fig:DoS}d)
is qualitatively the same as that of an FTI with a TI. As a result,
at such an interface both spin states propagate into the same direction.
This is consistent with the existence of chiral fermion modes at 
magnetic domain walls on the surface of 3D TIs \cite{Hasan}.

Another observation is that at the edges of an FTI only gapless states
with either clockwise or counterclockwise propagation direction can
exist depending on the direction of the FEF. The spin on a given edge
depends on whether the interface is with a TI/FTI or with an ordinary
insulator.

Local densities of states shown in Fig.~\ref{fig:DoS}a-d were calculated
for a Fermi energy $E_{F}=0$ in the center of the bulk gap. The energy
dependence of the transmission and reflection probability (back into
the opposite edge state) of an incoming electron is shown in Fig.~\ref{fig:DoS}e
for the situation in Fig.~\ref{fig:DoS}c. Here, reflection is perfect
for energies within the bulk gap. Analogously, for the situations
in Figs.~\ref{fig:DoS}a, b, and d transmission is perfect (not shown).
For energies outside the bulk gap the transmission and reflection
probabilities drop quickly, because an incoming electron is scattered
into bulk states then and it becomes unlikely that it ends up in one
of the edge states.

\subsection*{Devices}

The basic concept of all devices discussed in the following is the
bandstructure modification of thin films of TIs by local FEFs. These
could be either induced by ferromagnetic materials (FM) on top of
the TI via proximity effect or by doping with magnetic atoms. Using
the results for the propagation channels at different interfaces shown
in Fig.~\ref{fig:DoS}, we can now construct useful spintronic devices.

In our simulations the FTI areas have an edge length of $64-128$
atoms corresponding to $26.5-53\,\textrm{nm}$. However, the device
structures presented in the following in principle could be constructed
on an even smaller scale, because the functioning of the devices in
only limited by the spatial extent of the gapless edge states, which
can be of the order of 1~nm for large band gap materials \cite{Pauly}.

\subsubsection*{Spin current generator}

\begin{figure}
\includegraphics[width=1\columnwidth]{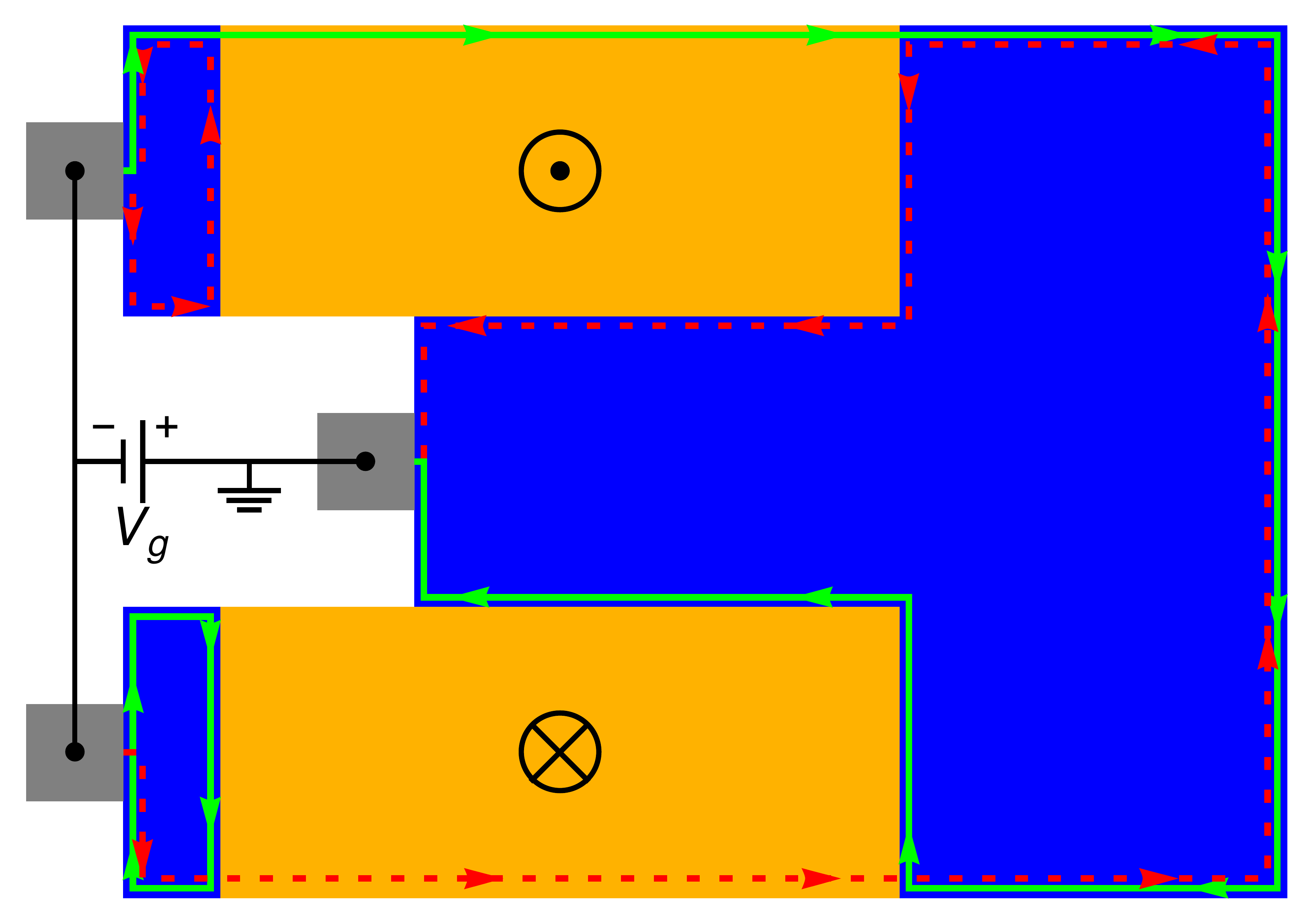}

\caption{\label{fig:Spin-current-generator}\textbf{Spin current generator.}
The spin current generator consists of two injecting and one extracting
metallic electrodes (gray) attached to a TI sheet (blue) with an applied
voltage $V_{g}$. Two FTI layers (orange) of opposite polarity prevent
direct current flow from the negative to the positive electrode, driving
spin currents along the right edge. Current flow in the two spin states
is indicated by green solid and red dashed arrows, respectively. The
black arrows represent the $z$-polarization of the FM layers (up
for positive, down for negative).}
\end{figure}

The first device is a spin current generator that creates pure spin
currents in a TI. Due to the locking of spin and propagation direction,
any charge transport in edge states results in a net spin transport
along the edge. To obtain a pure spin current without any net charge
transport we have to drive currents of equal magnitude in both directions
along the edge. This can be realized by the device shown in Fig.~\ref{fig:Spin-current-generator}.
By applying a voltage $V_{g}$ between the inner positive and the
two outer negative metallic contacts (gray), electrons will flow from
the negative contacts to the positive contact. We place two FTI areas
(orange) of opposite, fixed polarity as illustrated in Fig.~\ref{fig:Spin-current-generator}.
As discussed above, only electrons with one polarity can pass the
FTI area while the others are reflected. When the upper FTI is polarized
in positive $z$-direction, only those electrons with mainly spin-up
can pass and must propagate along the right edge, because the other
propagation direction is forbidden. The same holds for spin-down electrons
at the lower FTI with opposite polarity, resulting in a net spin transport
without a net charge transport along the right edge.

In principle, the small TI area (blue) between the metallic electrodes
and the FTI is not necessary for the functioning of the device. However,
as the FM layer should not overlap with the metallic contact, it should
be easier to prepare the device including this area.

We note that in this device the charge current is fully converted
into a spin current in principle, as two electron charges $2e$ flowing
from the two outer metallic contacts to the inner contact are converted
into a total spin flow of $p\hbar$ along the right edge. Here, $0\le p\le1$
is the spin polarization of the edge states. Reported values for $p$ typically
range between 0.3 and 0.9 depending on material (see Ref.~\onlinecite{Goette}
and references therein). 
In the present model we have $p=D/B\approx 0.35$, 
because the edge states have opposite spin for the 
two considered orbitals, partially compensating each other.
The conversion efficiency is given by
$\Theta=\frac{ej_{s}}{\hbar j_{c}}=0.5p$, where $j_{c}$ is the
charge and $j_{s}$ the spin current density. This can be compared
with the conversion efficiency of the spin Hall effect as quantified
by the spin Hall angle $\Theta_{SH}$. The largest known spin Hall
angles are currently of the order of 0.1 \cite{Hoffmann,Ralph}.

The conversion efficiency of the present device depends on temperature
as thermal excitation of edge state electrons into bulk states can
appear. This thermal effect can be kept small, if the Fermi energy
is arranged in the center of the bulk gap and materials with large
bulk gaps are chosen. For Bi$_{2}$Se$_{3}$ considered here the bulk
gap of 0.3~eV will allow operation of the device at room temperature
with high conversion efficiency.

We want to point out that spin-flip scattering at the edges, that could be
caused by magnetic impurities or by the Rashba effect due to the
coupling to a substrate for example,
is not going to affect the conversion efficiency of the
device. First of all in the FTI areas spin-flip scattering is forbidden
because the spin-up and spin-down edge channels are spatially separated.
Along the right edge spin-flip scattering is possible and will affect
the resistance of the device, if the length of the edge becomes larger
than the spin-flip mean free path, which in topological edge states
was reported to be of the order of 2~$\textrm{\ensuremath{\mu}m}$\cite{Koenig1,Koenig2}.
However, the conversion efficiency is not changed by spin-flip scattering:
consider a spin-flip scattering site at the right edge. If the scattering
rate for spin-up electrons scattered backwards into the spin-down
channel is $\gamma$, the scattering rate for spin-down electrons
scattered backwards into the spin-up channel will be the same due
to reciprocity. Thus, the loss in one channel due to backward spin-flip
scattering will be fully compensated for by the reverse process. As
a result, both spin current components will remain unchanged by spin-flip
scattering and the ratio between spin and charge current will stay
the same. (The voltage $V_{g}$ necessary to drive these currents
will increase, though).

\subsubsection*{Spin current detector}

\begin{figure}
\includegraphics[width=1\columnwidth]{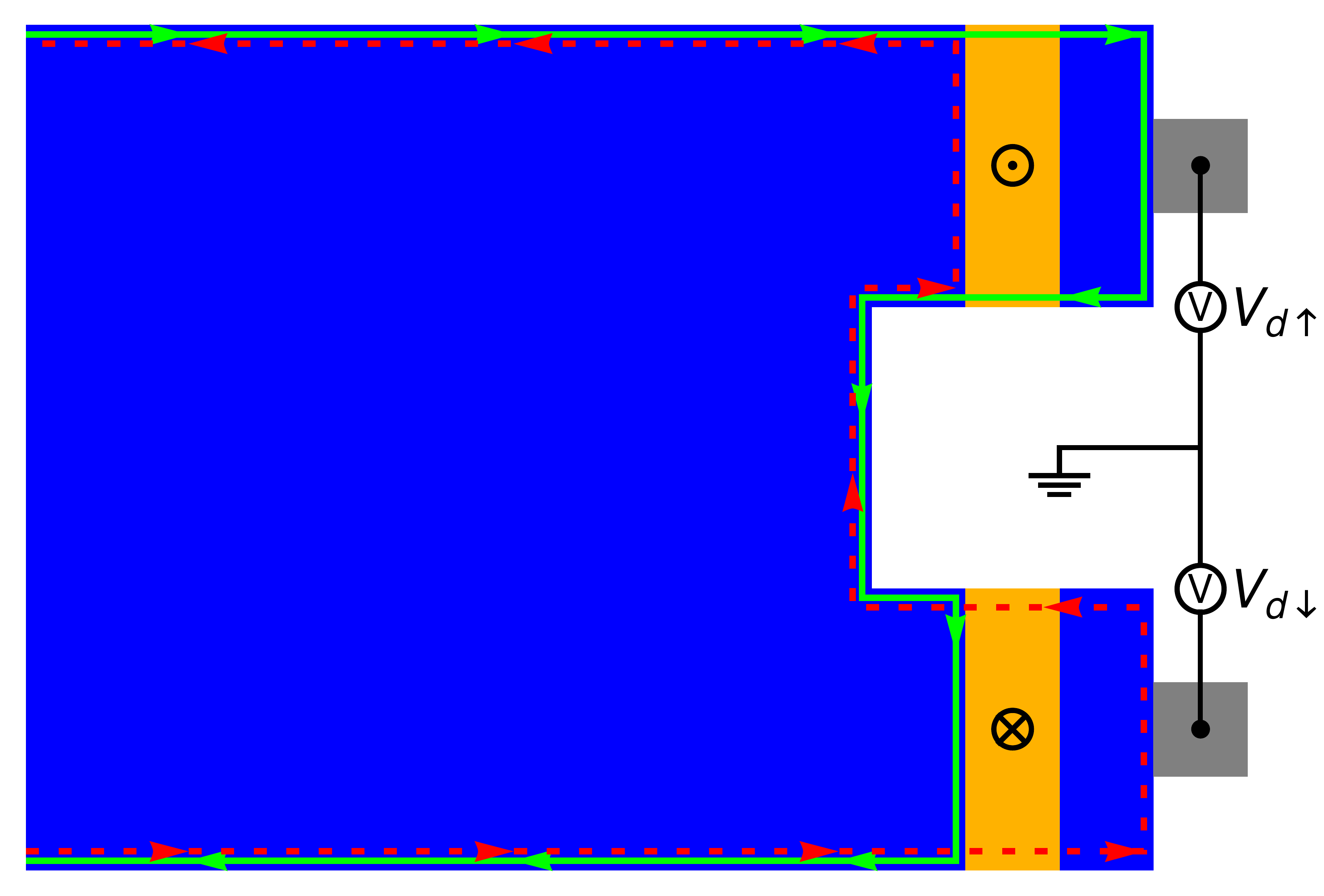}

\caption{\label{fig:Spin-current-detector}\textbf{Spin current detector.}
The spin current detector is basically an inverted generator which
consists of two FTI layers. This allows separate measurement of the
two spin polarized currents by blocking one while the other one can
pass. The resulting electrical currents can be measured as voltage
drops $V_{d\uparrow}$ and $V_{d\downarrow}$ with respect to the
common ground with the generator. In the case of a pure spin current,
these voltages are equal.}
\end{figure}

Because there is no instrument for direct measurement of spin currents,
the spin current needs to be transformed back into a charge current
in order to be detected. We do this by splitting the spin current
into its two counterpropagating parts and measure them individually.
This can be achieved by an inversion of the spin current generator,
as shown in Fig.~\ref{fig:Spin-current-detector}. The splitting
is done by two FM layers with opposite polarity. For positive $z$-magnetization,
only the spin-up current can pass the FTI and can be detected at the
upper electrode as a voltage drop $V_{d\uparrow}$ with respect to
the common ground with the generator. Such kind of voltage drops at
the edges of a 2D TI have been observed previously \cite{Bruene,Chang}.
Correspondingly, for negative $z$-magnetization, only the spin-down
part can pass the FTI and the resulting current can again be measured
as a voltage drop $V_{d\downarrow}$ at the lower electrode. For a
pure spin current, the two measured voltages will be equal, i.e. $V_{d\uparrow}=V_{d\downarrow}$.
With additional information on resistance and spin polarization of
the TI, the net transport of spin angular momentum can be calculated
from the measured voltage drop $V_{d}$.

\subsubsection*{Spin transistor}

\begin{figure}
\includegraphics[width=1\columnwidth]{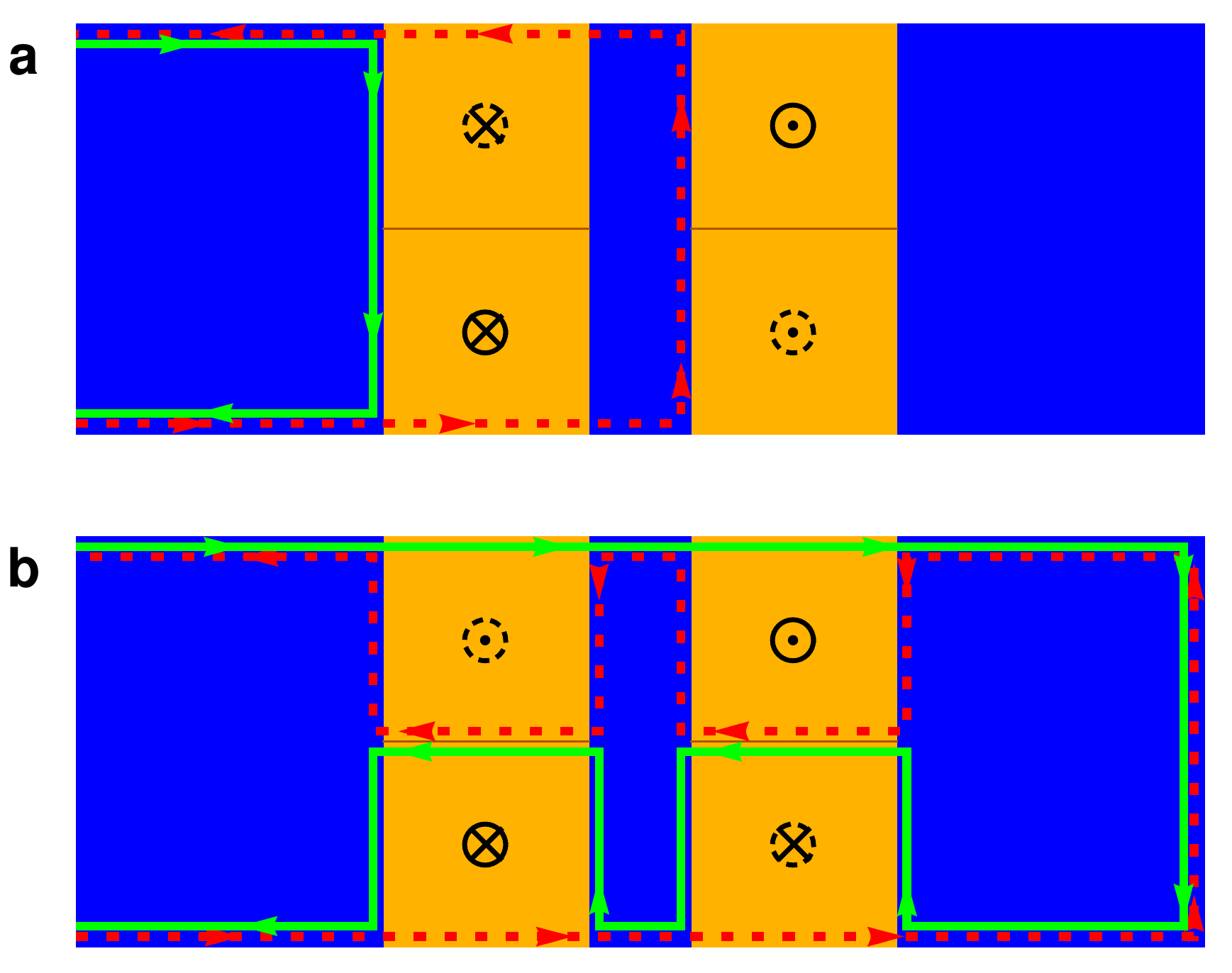}

\includegraphics[width=1\columnwidth]{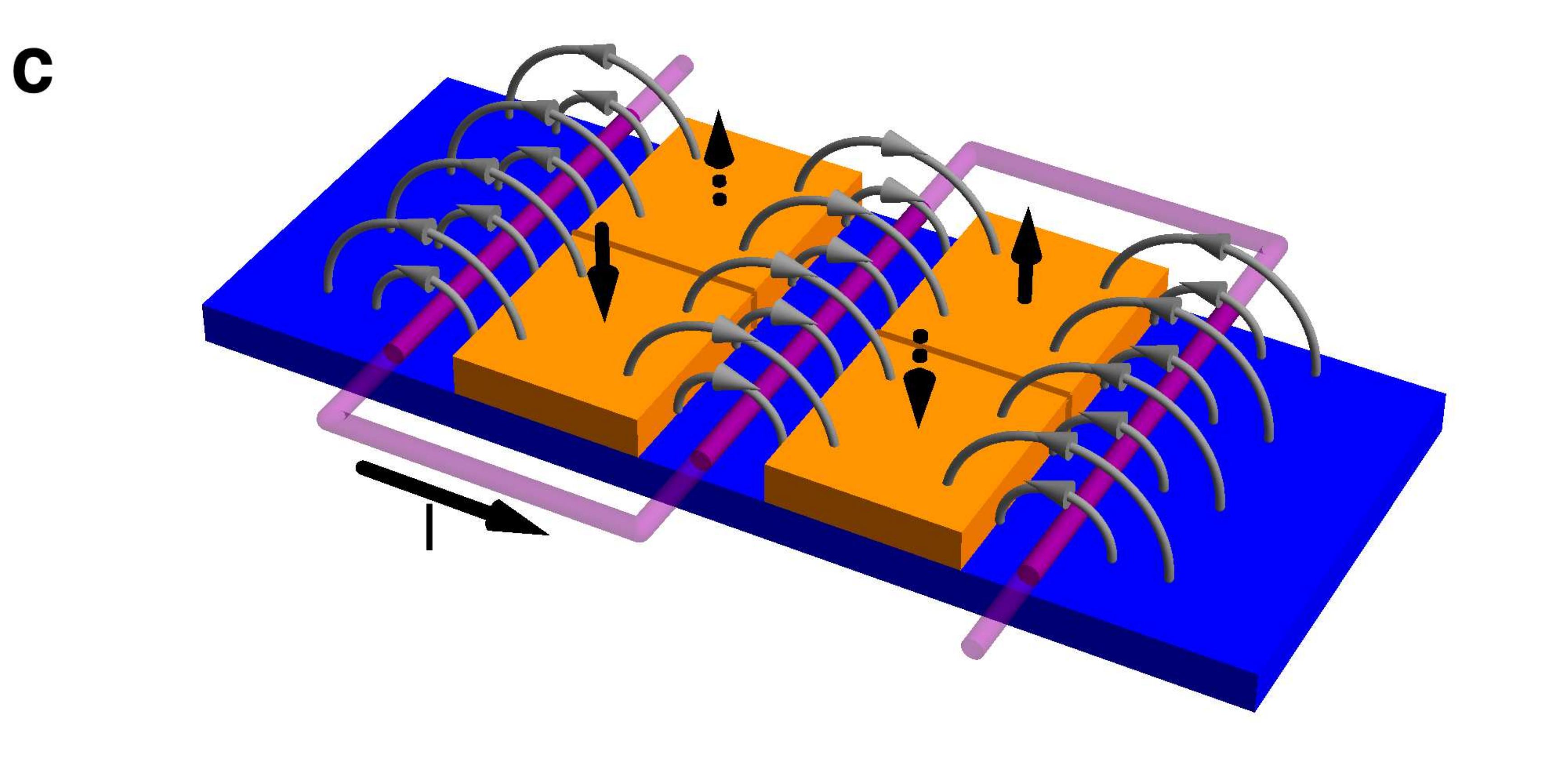}

\caption{\label{fig:Spin-transistor}\textbf{Spin transistor.} The spin transistor
device consists of four FTI domains from which two are switchable
(black dashed arrows). The other two FTI domains are fixed by exchange
bias pinning (black solid arrows). \textbf{a} and \textbf{b} show
the two configurations, where \textbf{a} blocks both spin states while
\textbf{b} allows them to pass. \textbf{c} shows a possibility to
switch the spin transistor. A current through the wire (purple) induces
a magnetic field (gray) with opposite polarity at the two switchable
magnetic domains. The loop makes the field stronger and homogeneous.}
\end{figure}

A spin transistor is a device that can either reflect or transmit
a pure spin current by switching between two configurations. In the
first configuration it should reflect all electrons and in the second
one it should let them pass, independent of their spin state. As a
single FTI layer can only block one spin direction while the other
one can pass, a spin transistor therefore requires a combination of
different FTI domains that can be switched individually. The most
simple device requires four magnetic domains in two blocks as shown
in Fig.~\ref{fig:Spin-transistor}. While the distance between the
two blocks is arbitrary, it is essential that the top and bottom domain
of each block are in direct contact in order to form a single domain
in the parallel orientation of both domains. In the first configuration
(``off'', Fig.~\ref{fig:Spin-transistor}a), both domains in a
block are oriented in parallel, while the orientation of the two blocks
is antiparallel. In this way, the left block (negative polarization)
reflects the spin-up part and the right block (positive polarization)
reflects the spin-down part of the spin current. The second configuration
(``on'', Fig.~\ref{fig:Spin-transistor}b) is reached by switching
two of the domains such that the two domains of each block have antiparallel
polarization. Here, we assume that both top domains have positive
polarization, while both bottom domains have negative polarization.
This allows free propagation from left to right along the upper edge
of the TI for spin-up electrons and along the lower edge for spin-down
electrons. It also opens a channel for the opposite direction at the
interface of the two FTI domains for both spin directions.

As the transmission and reflection rates are close to perfect, as
was discussed above (Fig.~\ref{fig:DoS}e), a high fidelity of this
spin transistor at room temperature can be expected, if the Fermi
level is chosen near the center of the bulk gap. Spin-flip scattering
due to magnetic impurities is possible whereever the two spin channels
are at the same location. However, similar as discussed above for
the spin current generator, due to reciprocity and the fact that there
are only two channels available, the scattering rate from the spin-up
to the spin-down channel is the same as vice versa. As a result, the
fidelity of this spin transistor is not affected by spin-flip scattering.

A possible set up to achieve switching between the two configurations
is shown in Fig.~\ref{fig:Spin-transistor}c. Two out of the four
magnetic domains should be fixed (solid arrows). This can be done
by exchange bias pinning with an antiferromagnetic material grown
on top of the domains\cite{Dieny}. The switching of the other two
domains (dashed arrows) could be performed by a magnetic field induced
by a current through a wire between the two FTI blocks as illustrated
in Fig.~\ref{fig:Spin-transistor}c. In this configuration the magnetic
field simultaneously switches both domains in opposite directions.
However, it may require high currents {(}of about $0.1\,\textrm{mA}-1\,\textrm{mA}${)}
to create a sufficiently strong magnetic field, potentially causing
heat problems. Alternatively, one could switch the domains separately
by a local external field, i.e. a write head. This has the disadvantage
of longer switching times. In addition, charge currents along the
right edge might appear during the time when only a single domain
is switched, if simultaneous switching is not possible.

As an alternative to the spin transistor device proposed here, one
could think of a device that possesses an insulating barrier that
can be turned on and off. An idea would be to use an FEF with in-plane
polarization, which opens a gap in the edge states \cite{RLChu2}.
However, the FEF would need to be disengageable for this purpose.
In the new class of topological crystalline insulators an electrical
field can open a gap in the edge states by breaking the underlying
mirror symmetry \cite{Liu}, which could also potentially be used
for the present purpose.

\section*{Discussion}

The recent advent of strong ferromagnetic topological insulators allows
separate control of the counterpropagating edge states in these materials.
We investigated gapless states in 2D TIs with local FEFs. A local
FEF perpendicular to the surface plane does not destroy one of the
two edge states, but pushes them inside the TI plane towards the edge
of the FEF. The edge states with opposite polarization remain basically
unchanged, however. This spatial separation of spin states motivates
the construction of spintronic devices. Here, we proposed devices
that allow the creation, switching and detection of pure spin currents.
Large bulk gaps in some TI materials may allow operation at room temperature,
because thermal excitation of edge state electrons into the bulk is
strongly suppressed by the bulk gap.
Note, that larger gaps also need larger exchange fields to drive the 
material into the QAH state, which might be a challenging requirement. 
Nevertheless, at low temperature, our devices should be feasible 
already in existing materials. Direct conversion of electrical
currents into spin currents makes our devices more efficient than
presently known methods of spin current generation. Eventually, the
efficiency depends only on the spin polarization of the TI material.
The devices in principle could be miniaturized down to the nanometer
scale, as the length scale is determined by the extent of the edge
states. The possibility to build all devices on a single TI sheet
avoids lattice mismatches at material interfaces, i.e. potential scattering
sites.

\section*{Methods}

\subsection*{Model}

In this work we investigate a thin film of Bi$_{2}$Se$_{3}$ as a
reference material, which belongs to a class of 3D TIs that can be
modelled by the two-orbital model Hamiltonian derived by Liu \textit{et
al.}\cite{Liu:PRB10}. If the film is thin enough, such that the top
and bottom surface states are gapped out, it can be reduced to an
effective 2D Hamiltonian using the quantum well approximation in perpendicular
direction\cite{RLChu,Paananen-PRB2013}. In the lattice regularized
version for a square lattice from Li \textit{et al.}\emph{\cite{Li:NPhys10}}
the Hamiltonian then reads 
\begin{equation}
H\left(\mathbf{k}\right)=\epsilon_{0}\left(\mathbf{k}\right)\mathbb{I}_{4\times4}+\underset{i=0}{\overset{2}{\sum}}m_{i}\left(\mathbf{k}\right)\Gamma^{i}\label{eq:hamiltonian}
\end{equation}
with $\epsilon\left(\mathbf{k}\right)=C+2D\left(2-\cos k_{x}-\cos k_{y}\right)$,
$m_{0}\left(\mathbf{k}\right)=M_{2D}-2B\left(2-\cos k_{x}-\cos k_{y}\right)$,
$m_{1}\left(\mathbf{k}\right)=A\sin k_{x}$, and $m_{2}\left(\mathbf{k}\right)=A\sin k_{y}$.
$\Gamma^{0,1,2}=\left(\tau_{z}\otimes\mathbb{I}_{2\times2},\tau_{x}\otimes\sigma_{x},\tau_{x}\otimes\sigma_{y}\right)$
are Dirac $\Gamma$ matrices in the basis of bonding and antibonding
$p_{z}$ orbitals, where the Pauli matrices $\tau_{i}$ and $\sigma_{i}$
operate in orbital and spin space, respectively. Following ref.~\onlinecite{Paananen-PRB2013}
we set $C=0$ and use an effective 2D parameter $M_{2D}=0.17\,\textrm{eV}$
corresponding to a film thickness of $3\textrm{nm}$. The remaining
parameters are derived from Zhang \textit{et al.}\cite{Zhang:NPhys09}
with an effective lattice constant $a=\left(\sqrt{3}/2\right)^{1/2}\cdot4.14\,\text{\AA}$:
$A=1.06\,\textrm{eV}$, $B=3.81\,\textrm{eV}$ and $D=1.32\,\textrm{eV}$.
The prefactor $\left(\sqrt{3}/2\right)^{1/2}$ to the real lattice
constant given in refs.~\onlinecite{Okamoto,Lind,Zhang:APL09}
is chosen such that the area of the first Brillouin zone of the square
lattice used here equals that of the actual hexagonal lattice.

Local FEFs are modelled by locally adding a Zeeman field of strength
$V_{z}$ in $z$-direction (perpendicular to the surface plane) 
\begin{equation}
H_{Z}=V_{z}\mathbb{I}_{2\times2}\otimes\sigma_{z}
\end{equation}

\subsection*{Eigenstates and wave-packets}

To calculate the propagation of edge state electrons and obtain their
transmission and reflection in our inhomogeneous devices, we use numerical
quantum transport calculations in analogy to the work by Krueckl and
Richter \cite{Krueckl}. In order to construct an incoming wave-packet
for the time evolution, we first consider an infinite strip without
FEF. We Fourier transform equation~\eqref{eq:hamiltonian} onto its
spatial lattice in the direction perpendicular to the interface, e.g.
in $y$-direction. In this case the momentum component parallel to
the interface remains a good quantum number and allows calculation
of all eigenstates and bandstructure for discrete momenta $k_{x}$
by exact numerical diagonalization. Results of such calculations including
an FEF are shown in Fig.~\ref{fig:Dispersion}b-e. From these eigenstates,
we choose those with energy inside the bulk gap and sort them into
four groups (denoted by $\nu$) selected by the sign of their group
velocity $v=\frac{1}{\hbar}\frac{\partial E\left(k_{x}\right)}{\partial k_{x}}$
and their spin polarization. The states $\chi_{\nu}\left(y,k_{x}\right)$
of each group are weighted by a Gaussian distribution 
\[
\eta_{\nu}\left(k_{x}\right)=\frac{1}{\left(2\pi d^{2}\right)^{1/4}}e^{-\frac{\left(k_{x}-k_{0}\right)^{2}}{4d^{2}}}
\]
avoiding a sharp momentum cut-off at the ends of the bulk gap. This
Gaussian distribution is located around the mean value $k_{0}$ of
these $k_{x}$.
Then the weighted states are Fourier transformed in $k_{x}$ to construct
wave-packets located around a certain starting position $x_{0}$ 
\[
\Phi_{\nu}\left(x,y\right)=\frac{1}{\sqrt{2\pi}}\underset{k_{x}}{\sum}\eta_{\nu}\left(k_{x}\right)\chi_{\nu}\left(y,k_{x}\right)e^{ik_{x}\left(x-x_{0}\right)}\Delta k_{x}
\]
The sum runs over all $k_{x}$, for which $\chi_{\nu}\left(y,k_{x}\right)$
is defined, in steps of $\Delta k_{x}$.

\subsection*{Time evolution}

For the time evolution of a wave-packet, equation~\eqref{eq:hamiltonian}
is Fourier transformed in both directions onto a lattice of size $1024\times128$.
Periodic boundary conditions are applied in $x$-direction. Local
FTI areas have a size of $128\times64$ or $128\times128$, as illustrated
by the orange areas in Fig.~\ref{fig:DoS}. The time-evolution of
a wave-packet is then calculated by numerically applying the time-evolution
operator $U\left(x,y,t\right)=\exp\left(-iH\left(x,y\right)t/\hbar\right)$
to a starting wave-packet initially located close to one of the corners
of the lattice.

Following the method of Krueckl and Richter \cite{Krueckl}, from
the time dependent wave function the energy dependent transmission
probabilities 
\[
\left|S_{\beta,\alpha}\right|^{2}\left(E\right)=\frac{v^{2}\left(E\right)}{\left(2\pi\right)^{4}\eta^{4}\left(E\right)}\left|\underset{t}{\sum}C_{\beta,\alpha}\left(t\right)e^{iEt/\hbar}\Delta t\right|^{2}
\]
are calculated via Fourier transformation in time of the time dependent
overlap $C_{\beta,\alpha}\left(t\right)$ of the propagating wave-packet
(denoted by $\alpha$) with four exit wave-packets (denoted by $\beta$)
located close to the corners of the lattice. A prefactor removes dependencies
on the density of states and the Gaussian weighting factor. The energy
dependent quantities $\eta\left(E\right)$ and the absolute value
of the group velocity 
\[
v\left(E\right)=\frac{A}{\hbar}\sqrt{1-\frac{D^{2}}{B^{2}}}\sqrt{1-\frac{\left(B\cdot E-D\cdot M_{\textrm{2D}}\right)^{2}}{A^{2}\left(B^{2}-D^{2}\right)}}
\]
(obtained from the dispersion relation equation.~(95) in ref.~\onlinecite{PGGD})
do not depend on the index $\nu$. The local densities of states shown
in Fig.~\ref{fig:DoS} are proportional to the absolute value of
the time-integrated propagating wave-packet 
\[
\left|\psi_{E_{F}}\right|^{2}\left(x,y\right)\propto\left|\underset{t}{\sum}U\left(x,y,t\right)\Phi_{\nu}\left(x,y\right)e^{iE_{F}t/\hbar}\Delta t\right|^{2}
\]
where the factor $\exp\left(iE_{F}t/\hbar\right)$ allows to select
the Fermi energy $E_{F}$. For more details see Krueckl and Richter\cite{Krueckl}
and the corresponding supplementary material.

\section*{Acknowledgements}

Financial support from the DFG via SPP 1666 ``Topological Insulators''
is gratefully acknowledged. We acknowledge support for the Article 
Processing Charge by the Deutsche Forschungsgemeinschaft and the 
Open Access Publication Funds of Bielefeld University Library.

\section*{Author contributions}

M.G. and M.J. developed the quantum transport code. M.G. devised the
devices, analyzed the numerical results, and prepared the figures.
M.G. and T.D. wrote the manuscript. T.D. conceived and supervised
the project.

\section*{Competing interests}

The authors declare no competing financial interests.

\end{document}